\documentclass{article}
\usepackage[english]{babel}
\usepackage{amsmath,amssymb,graphicx,bbm,hyperref,xcolor}

\begin{document}

\title{A $Sim (2)$   invariant  dimensional regularization}

\author{J. Alfaro \\
	Facultad de F\'\i sica, Pontificia Universidad Cat\'olica de Chile,\\
	Casilla 306, Santiago 22, Chile.\\
	jalfaro@uc.cl}

\maketitle

\begin{abstract}
  We introduce a $Sim (2)$ invariant dimensional regularization of
  loop integrals. Then we can compute the one loop quantum corrections to the
  photon self energy, electron self energy and vertex in the Electrodynamics
  sector of the Very Special Relativity Standard Model(VSRSM). 
\end{abstract}

The Weinberg-Salam model(SM) is a very successfull description of Nature, that
is being verified at the LHC with a great precission. Moreover, until now,
neither new particles nor new interactions have been discovered at the
LHC{\cite{w2}}. This cannot be the whole story, though. The SM assumes that
the neutrino is a massless particle, whereas we know that the neutrino is
massive in order to describe the observed neutrino
oscillations{\cite{Langacker}}

If we assume that Lorentz's is an exact symmetry of Nature, we have to
introduce new particles and interactions in order to give masses to the
observed neutrinos through, for instance, the seesaw
mechanism{\cite{mohapatra}}.

A new possibility to have a massive neutrino arises in Very Special
Relativity(VSR){\cite{CG1}}. \ Instead of the 6 parameter Lorentz group, a 4
parameters subgroup($Sim(2)$) is assumed to be the symmetry of
Nature. $Sim(2)$ transformations change a fixed null four vector
$n_{\mu}$ at most by a scale factor, so ratios of scalar quantities containing
the same number of $n_{\mu}$ in the numerator as in the denominator are
$Sim(2)$ invariant, although \ they are not Lorentz invariant. In
this way it is possible to write a VSR mass term for left handed
neutrinos{\cite{CG2}}.

Recently, we have proposed the SM with VSR{\cite{ja1}} (VSRSM).It contains the
same particles and interactions as \ the SM, but neutrinos can have a VSR mass
without lepton number violation. Since the electron and the electron neutrino
form a $SU(2)_{L}$ doublet, the VSR neutrino mass term will modify
the QED of the electron.

A main obstacle in exploring the loop corrections in the VSRSM is the
non-existence of a gauge invariant regulator that preserve the $Sim(2)$ symmetry of the model.

In this letter, we define an appropriate regulator, based on the calculation
of integrals using the Mandelstam-Leibbrandt(ML){\cite{Mandelstam}}
{\cite{Leibbrandt}}prescription introduced in {\cite{alfaroML}}.We want to
emphasize that our method directly lead to the ML prescription, the only one
compatible with canonical quantum field theory{\cite{Soldati}}.The regulator
preserve gauge invariance, a property inherited from the ML prescription, as
well as the $Sim(2)$ symmetry.

Then we proceed to compute \ one loop corrections. We find the divergent and \
finite part of the vacuum polarization and electron self energy. Moreover we
compute the leading correction to the standard QED result for the anomalous
magnetic model of the electron.

We want to emphasize that meanwhile no new particles or interactions are
discovered at the LHC or elsewhere, we have to consider the VSRSM as a very
strong candidate to describe weak and electromagnetic interactions. It
contains all the predictions of the SM plus neutrino masses and neutrino
oscillations. It is renormalizable(as we show explicitly in this letter) and
unitarity of the $S$ metric is preserved. If future experiments validates the
predictions of the model, it would be the first evidence of Lorentz Symmetry
violation.

\section{Mandelstam-Leibbrandt(ML) prescription from a hidden symmetry}

\ \ In this section we review the results of  {\cite{alfaroML}}.

Let us compute the following simple integral:
\[ A_{\mu} = \int dp \frac{f (p^{2} ) p_{\mu}}{(n \cdot p)} \]
where $f$ is an arbitrary function.$dp$ is the integration measure in $d$
dimensional space and $n_{\mu}$ is a fixed null vector($(n \cdot n) =0$). This
integral is infrared divergent when $(n \cdot p) =0$.

The ML is:
\begin{equation}
  \frac{1}{(n \cdot p)} = \lim_{\varepsilon \rightarrow 0}   \frac{(p \cdot
  \bar{n} )}{(n \cdot p)  (p \cdot \bar{n} ) +i \varepsilon} \label{ml}
\end{equation}
where $\bar{n}_{\mu}$ is a new null vector with the property $(n \cdot \bar{n}
) =1$.

To compute $A_{\mu}$ we must know what $f$ is , provide an specific form of \
$n_{\mu}$ and $\bar{n}_{\mu}$, and evaluate the residues of all poles of \
$\frac{f (p^{2} )}{(n \cdot p)}$ in the $p_{0}$ complex plane, a difficult task for an arbitrary $f$.

Instead we want to point out the following symmetry:
\begin{equation}
  n_{\mu} \rightarrow \lambda n_{\mu} , \bar{n}_{\mu} \rightarrow \lambda^{-1}
  \bar{n}_{\mu} , \lambda \neq 0, \lambda \varepsilon R \label{symmetry}
\end{equation}
It preserves the definitions of $n_{\mu}$ and $\bar{n}_{\mu}$:
\begin{eqnarray*}
  0= (n \cdot n) \rightarrow \lambda^{2}  (n \cdot n) =0 &  & \\
  0= ( \bar{n} \cdot \bar{n} ) \rightarrow \lambda^{-2}  ( \bar{n} \cdot
  \bar{n} ) =0 &  & \\
  1= (n \cdot \bar{n} ) \rightarrow (n \cdot \bar{n} ) =1 &  & 
\end{eqnarray*}
We see from (\ref{ml}) that:
\[ \frac{1}{(n \cdot p)} \rightarrow \frac{1}{(n \cdot p)} \lambda^{-1} \]
Now we compute $A_{\mu}$, based on its symmetries. It is a Lorentz vector
which scales under (\ref{symmetry}) as $\lambda^{-1}$. The only Lorentz
vectors we have available in this case are $n_{\mu}$ and $\bar{n}_{\mu}$. But
(\ref{symmetry}) forbids $n_{\mu}$. That is:
\[ A_{\mu} =a \bar{n}_{\mu} \]
Multiply by \ $n_{\mu}$ to find $(A \cdot n) =a$. Thus $a= \int dpf (p^{2} )$.
Finally:
\[ \int dp \frac{f (p^{2} ) p_{\mu}}{(n \cdot p)} = \bar{n}_{\mu}  \int dpf
   (p^{2} ) \]
By the same arguments, we can compute the generic integral:
\begin{equation}
  A= \int dp \frac{F (p^{2} ,p \cdot q)}{(n \cdot p)} = ( \bar{n} \cdot q) f
  (q^{2} ,(n \cdot q)( \bar{n} \cdot q)) \label{generic}
\end{equation}
$q_{\mu}$ is an external momentum, a Lorentz vector. $F$ is an arbitrary
function. The last relation follows from (\ref{symmetry}), for a certain $f$
we will find in the following.

Taking the partial derivative respect to $q_{\mu}$ in both sides of
(\ref{generic}), we obtain that
\begin{eqnarray}
  \frac{\partial A}{\partial q^{\mu}} n_{\mu} = \int dpF_{,u} = & g (x) = & 
  \nonumber\\
  f (x,y) +2y \frac{\partial}{\partial x} f (x,y) +y \frac{\partial}{\partial
  y} f (x,y) &  &  \label{pdi1}
\end{eqnarray}
We defined $u=p \cdot q$,$x=q^{2}$,$y= (n \cdot q)  ( \bar{n} \cdot q)$.
$()_{,u}$ means derivative respects to $u$.

Assuming that the solution and its partial derivatives are finite in the
neighborhood of $y=0$, it follows from the equation that $f (x,0) =g (x)$.
That is the partial differential equation has a unique regular solution.

Now we apply this result to compute integrals that appear in gauge theory
loops:
\[ \int dp \frac{1}{[p^{2} +2p \cdot q-m^{2} ]^{a}}  \frac{1}{(n \cdot p)} = (
   \bar{n} \cdot q) f (x,y) \]
In this case
\[ g (x) =-2a \int dp \frac{1}{[p^{2} -x-m^{2} ]^{a+1}} \]
The unique regular solution of (\ref{pdi1}) is:
\begin{eqnarray*}
  f ( x,y ) =- \frac{1}{y}  \left\{ \int dp[p^{2} -x-m^{2} ]^{-a} - \int
  dp[p^{2} -x+2y-m^{2} ]^{-a} \right\} &  & 
\end{eqnarray*}
We can check \ that $f (x,0) =-2a \int dp [p^{2} -x-m^{2} ]^{-a-1} =g (x)$.

In the same way we can compute the whole family of loop integrals:
\[ \int dp \frac{1}{[p^{2} +2p \cdot q-m^{2} ]^{a}}  \frac{1}{((n \cdot
   p))^{b}} = ( \bar{n} \cdot q)^{b} (-2)^{b} \frac{\Gamma (a+b)}{\Gamma (a)
   \Gamma (b)}  \int_{0}^{1} dtt^{b-1}  \int dp [p^{2} -q^{2} +2n.q \bar{n}
   .qt-m^{2} ]^{-a-b} \]
Using dimensional regularization, we obtain:
\begin{eqnarray}
  \int dp \frac{1}{[p^{2} +2p \cdot q-m^{2} ]^{a}}  \frac{1}{((n \cdot
  p))^{b}} = &  &  \nonumber\\
  (-1)^{a+b} i ( \pi )^{\omega} (-2)^{b} \frac{\Gamma (a+b- \omega )}{\Gamma
  (a) \Gamma (b)}  ( \bar{n} \cdot q)^{b}  \int_{0}^{1} dtt^{b-1} 
  \frac{1}{(m^{2} +q^{2} -2(n \cdot q)( \bar{n} \cdot q)t)^{a+b- \omega}} ,
  \omega =d/2 &  &  \label{I1}
\end{eqnarray}

\section{$Sim(2)$ invariant regulator}

The prescription to regularize the infrared divergences that we have reviewed
in chapter 1, always produces finite results depending on two fixed null
vectors $\bar{n}_{\mu} ,n_{\mu}$. Moreover it preserves gauge invariance
because it respects the shift symmetry of the loop integral $\int dpf (p_{\mu}
) = \int dpf (p_{\mu} +q_{\mu} )$ for arbitrary $q_{\mu}$. However ML does not
respect $Sim (2)$ symmetry of VSRSM. Below we show how to remedy this.

We start from the ML result for the integral(\ref{I1}).

We trade $\bar{n}_{\mu}$ by $q_{\mu}$. i.e. $\bar{n}_{\mu} =an_{\mu}
+bq_{\mu}${\footnote{This is the more general form for $\bar{n}_{\mu}$
compatible with reality, right scaling under (\ref{symmetry}) and
$\bar{n}_{\mu}$ dimensionless.

For instance in $d=3$ we must have $\bar{n}_{\mu} =a \frac{q^{2} n_{\mu}}{(
n.q )^{2}} +b \frac{q_{\mu}}{n.q} +c  \varepsilon_{\mu \nu \lambda}
\frac{n_{\nu}}{( n.q )^{2}} q_{\lambda} \sqrt{q^{2}}$ with $a,b,c$ pure
numbers. This fails to be real for $q^{2} <0$.}}. From the conditions:
$\bar{n} . \bar{n} =0$,$  \bar{n} .n=1$ we get $\bar{n}_{\mu} =-
\frac{q^{2}}{2 (n.q)^{2}} n_{\mu} + \frac{q_{\mu}}{n.q}$. Moreover, we see
that $\bar{n}_{\mu}$ satisfies the scaling (\ref{symmetry}) and is real for
any value of $q^{2}$ in Minkowsky space. So, all the conditions to apply the
procedure reviewed in section 1 are satisfied. Therefore,,
\begin{eqnarray}
  \int d p \frac{1}{[ p^{2} +2p.q-m^{2} ]^{a}} \frac{1}{( n.p )^{b}} = &  & 
  \label{simI1}\\
  ( -1 )^{a+b} i ( \pi )^{\omega} ( -2 )^{b} \frac{\Gamma ( a+b- \omega
  )}{\Gamma ( a ) \Gamma ( b )} \left( \frac{q^{2}}{2n \cdot q} \right)^{b}
  \int_{0}^{^{1}} d t t^{b-1} \frac{1}{( m^{2} +q^{2} (  1-t ) )^{a+b-
  \omega}} , & \omega =d/2 &  \nonumber
\end{eqnarray}
Notice that now (\ref{simI1}) respects the $Sim (2)$ invariance of the
original integral. The same procedure can be applied to other integrals found
in {\cite{alfaroML}}. Notice that first we keep $\bar{n}$ fixed, derive
(\ref{I1}) with respect to $q_{\mu}$ as many times as necessary and then
replace $\bar{n}_{\mu} =- \frac{q^{2}}{2 (n.q)^{2}} n_{\mu} +
\frac{q_{\mu}}{n.q}$. The rationale for this prescription derives from the
observation that we could compute the integral with whatever power of
$p_{\mu}$ in the numerator using Cauchy theorem of residues in $p_{0}$ complex
plane. In this way it doesn't matter whether $\bar{n}_{\mu}$ depends on
$q_{\mu}$ or not.

Once we have obtained (\ref{simI1}), we notice that it provides a unique analytic continuation of the integral from $b<0$ to $b>0$.
Since for $b<0$ we do not need an infrared regulator, we can compute the integral using standard dimensional regularization. By integration by parts in the integral over $t$,we can check that (\ref{simI1}) gives the right answer for $b<0$.

\section{The model}

The leptonic sector of VSRSM consists of three $SU (2)$ doublets $L_{a} =
\left( \begin{array}{c}
  \nu^{0}_{aL}\\
  e^{0}_{aL}
\end{array} \right)$, where $\nu^{0}_{aL} = \frac{1}{2}  (1- \gamma_{5} )
\nu^{0}_{a}$ and $e^{0}_{aL} = \frac{1}{2}  (1- \gamma_{5} ) e^{0}_{a}$, and
three $SU (2)$ singlet $R_{a} =e^{0}_{aR} = \frac{1}{2}  (1+ \gamma_{5} )
e^{0}_{n}$. We assume that there is no right-handed neutrino. The index $a$
represent the different families and the index $0$ say that the fermionic
fields are the physical fields before breaking the symmetry of the vacuum.

In this letter we restrict ourselves to the electron family.

$m$ is the VSR mass of both electron and neutrino. 

After spontaneous symmetry breaking(SSB), the electron adquires a mass term
$M= \frac{G_{e} v}{\sqrt{2}}$, where $G_{e}$ is the electron Yukawa coupling
and $v$ is the VEV of the Higgs. Please see equation (52) of {\cite{ja1}}. The neutrino mass is not affected by SSB:$M_{\nu_{e}} =m$.

Restricting the VSRSM after SSB to the interactions between photon and
electron alone,  we get the VSR QED action.$\psi$ is the electron field.
$A_{\mu}$ is the photon field. We use the Feynman gauge.
\begin{eqnarray*}
  \mathcal{L}= \bar{\psi} \left( i \left( \not{D} + \frac{1}{2} \not{n} m^{2}
  (n \cdot D)^{-1} \right) -M \right) \psi - \frac{1}{4} F_{\mu \nu} F^{\mu
  \nu} - \frac{( \partial_{\mu} A_{\mu} )^{2}}{4} &  & \\
  D_{\mu} = \partial_{\mu} -i e A_{\mu} , & F_{\mu \nu} = \partial_{\mu}
  A_{\nu} - \partial_{\nu} A_{\mu} & 
\end{eqnarray*}
We see that the electron mass is $M_{e} = \sqrt{M^{2} +m^{2}}$, where $m$ is
the electron neutrino mass.

\subsection{Feynman rules}
To draw the Feynman graphs we used \cite{ellis}
\begin{figure}[h]
	\includegraphics[scale=0.4]{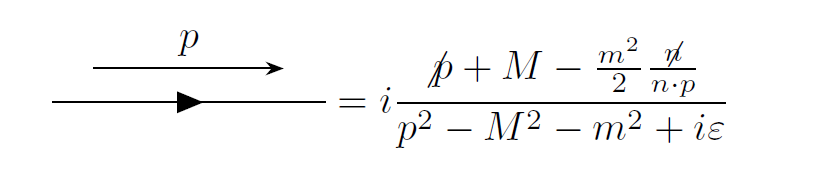}
	\includegraphics[scale=0.4]{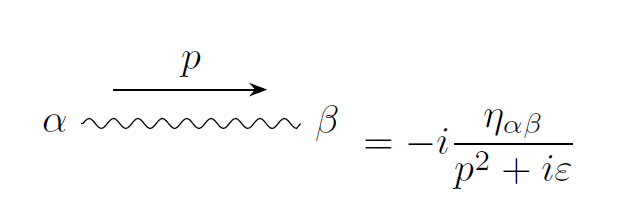}
	\includegraphics[scale=0.4]{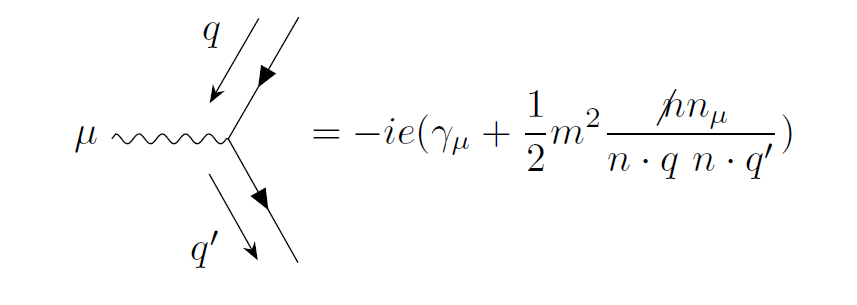}
	\includegraphics[scale=0.4]{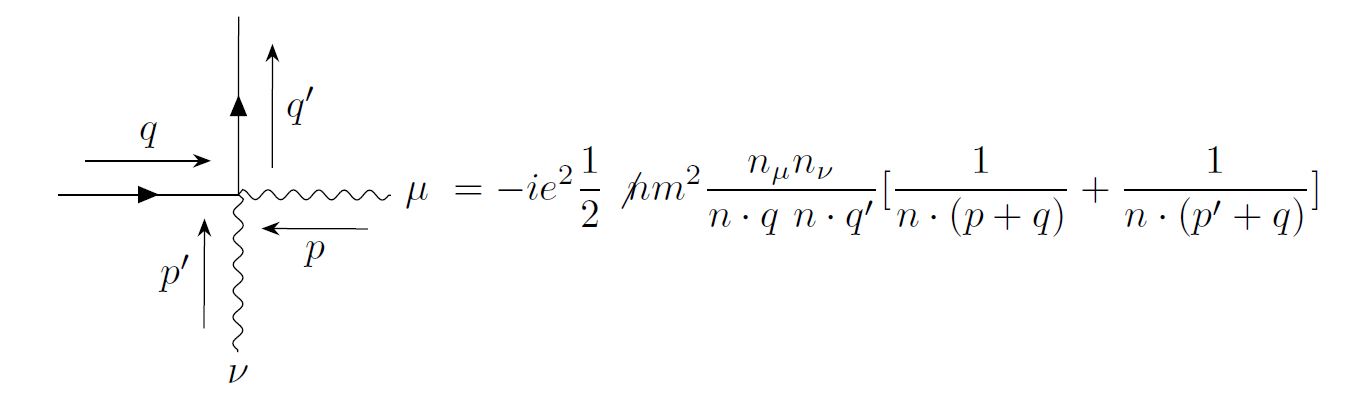}
	\caption{Feynman rules for one loop computations:electron propagator,photon propagator,$A_\mu ee$ and $A_\mu A_\nu ee$ vertex.}
	\label{Fig: Fey rules}
\end{figure}	
In the following sections, we have used extensively the program FORM {\cite{form}}.

\section{Photon Self Energy in VSRSM}

In this section we present the computation of the photon self-energy. In VSRSM
it is given by two graphs:

\begin{figure}[h]
	\centering
	\includegraphics[width=0.2\textwidth]{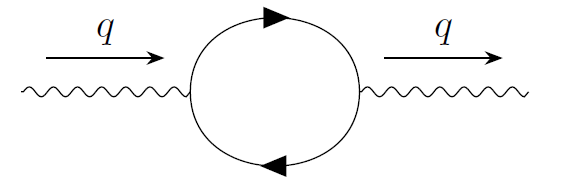}
	\includegraphics[width=0.2\textwidth]{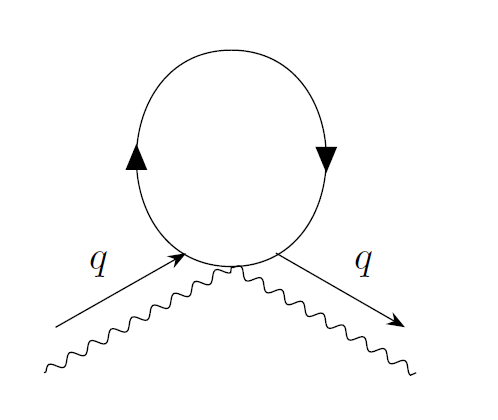}	
	\caption{Vacuum polarization one loop graphs}
	\label{Fig:vp}
\end{figure}

Applying the $Sim(2)$ invariant regulator to the addition of the graphs of Figure (\ref{Fig:vp}), and after a long calculation,
we get:
\begin{equation}
  i \Pi_{\mu \nu} =A ( \eta_{\mu \nu} q^{2} -q_{\mu} q_{\nu} ) +B \left(
  -q^{2}  \frac{n_{\mu} n_{\nu}}{(n.q)^{2}} + \frac{n_{\mu} q_{\nu} +n_{\nu}
  q_{\mu}}{n.q} - \eta_{\mu \nu} \right)
\end{equation}
with
\begin{eqnarray}
  A=(-ie)^{2} \frac{i}{(4 \pi )^{\omega}}  \int_{0}^{1} dx \Gamma (2- \omega )
  \frac{8x (1-x)}{(M_{e}^{2} -(1-x)xq^{2} )^{2- \omega}} &  &  \nonumber\\
  B=-m^{2} i {\color{red} } \frac{e^{2}}{4 \pi^{2}} \int_{0}^{1} \frac{d
  x}{{(1-x)}}   \log \left[ 1- \frac{q^{2} ( 1-x )^{2}}{M_{e}^{2}
  -q^{2} ( 1-x ) x} \right] &  & 
\end{eqnarray}
Here $-e$ is the electron electric charge, $m$ the electron neutrino mass
and $M_{e}$ is the electron mass.$q_{\mu}$ is the virtual photon momentum.

We first notice that $q^{\mu} \Pi_{\mu \nu} =0$ as required by $U (1)$ gauge
invariance of the photon field. It is obtained by a straightforward application of the regularized integrals of  . Moreover $B (q^{2} =0) =0$,therefore the
photon remains massless. Also the photon wave function divergence is the same
as in QED.

\section{Electron Self Energy in VSRSM}

\begin{figure}[h]
	\centering
	\includegraphics[width=0.2\textwidth]{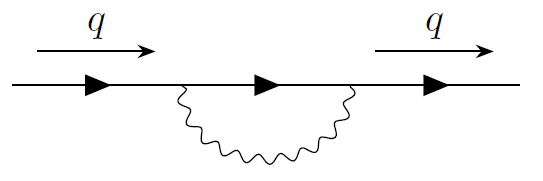}
	\includegraphics[width=0.2\textwidth]{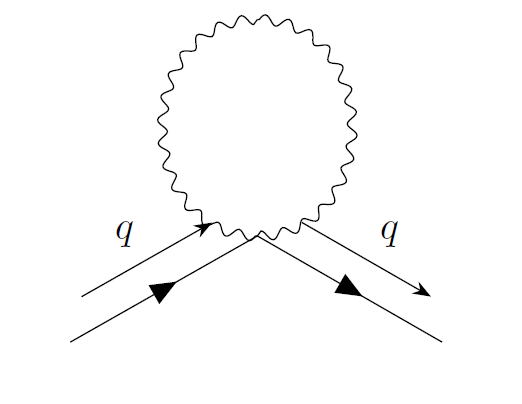}	
	\caption{Electron self energy one loop graphs. The second graph vanishes in Feynman gauge.}
	\label{Fig:ese}
\end{figure}

Here we calculate the electron self-energy. Again we have two graphs contributing to the 2-proper vertex.
See Figure(\ref{Fig:ese}).

\begin{equation}
  -i  \Sigma ( q ) =C \frac{\not{n}}{n.q} +D \not{q} +E 
\end{equation}
with:
\begin{eqnarray}
  C= ( -i e )^{2} m^{2} [  \frac{i}{16 \pi^{2}} \int_{0}^{1} d x (
  1-x )^{-1} \ln \left( 1+ \frac{q^{2} ( 1-x )}{( M_{e}^{2} -q^{2} -i
  \varepsilon )} \right) + &  &  \nonumber\\
  2i  ( 4 \pi )^{- \omega} \int_{0}^{1} d x \frac{\Gamma ( 2- \omega )}{[
  \mu^{2} x-x ( 1-x ) q^{2} + ( M^{2} +m^{2} ) ( 1-x ) -i \varepsilon ]^{2-
  \omega}} ] , &  &  \nonumber\\
  D=-2 ( -i e )^{2} ( \omega -1 ) i  ( 4 \pi )^{- \omega} \int_{0}^{1} d x
  \frac{\Gamma ( 2- \omega ) x}{[ \mu^{2} x-x ( 1-x ) q^{2} + ( M^{2} +m^{2} )
  ( 1-x ) -i \varepsilon ]^{2- \omega}} , &  &  \nonumber\\
  E= ( -i e )^{2} 2 \omega  M i  ( 4 \pi )^{- \omega} \int_{0}^{1} d x
  \frac{\Gamma ( 2- \omega )}{[ \mu^{2} x-x ( 1-x ) q^{2} + ( M^{2} +m^{2} ) (
  1-x ) ]^{2- \omega}} &  & 
\end{eqnarray}

\section{Electron-Electron-Photon Proper vertex $\Gamma^{\mu} ( p+q,p
)$}

In this subsection we discuss the 3 points proper vertex and verify the
Ward-Takahashi identity. This is an important test of the gauge invariance of
the regulator. The one loop contribution to $\Gamma^{\mu} ( p' = p+q,p )$ consists of
the addition of 3 graphs(Figure(\ref{Fig:3-vertex})):
\begin{figure}[h]
	\centering
	\includegraphics[width=0.2\textwidth]{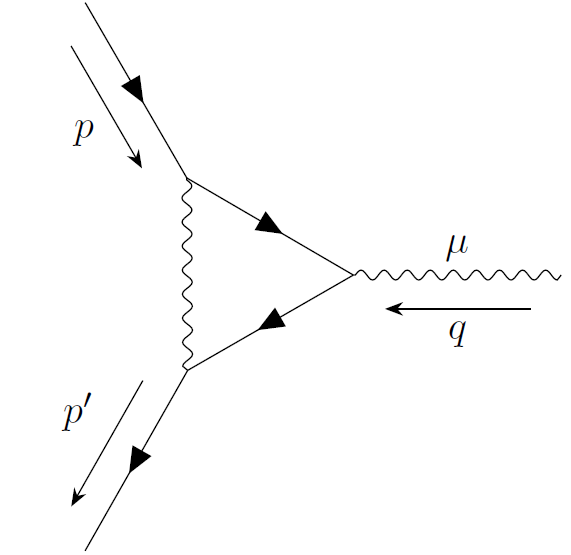}
	\includegraphics[width=0.2\textwidth]{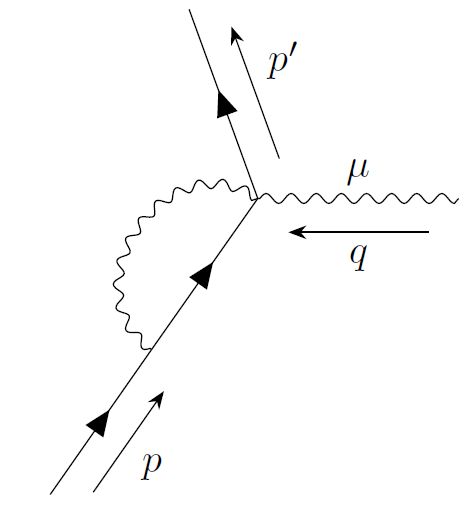}
	\includegraphics[width=0.2\textwidth]{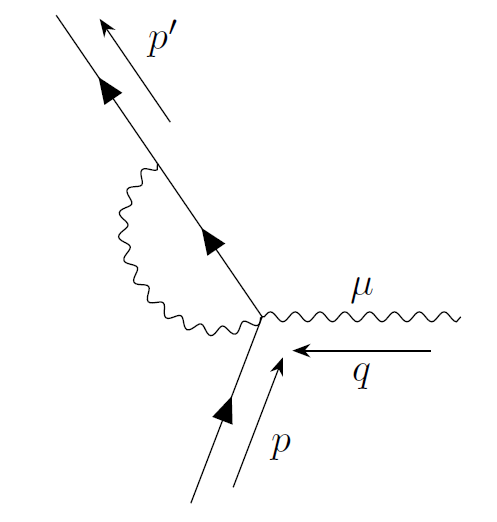}
	\caption{One loop contribution to the 3 points proper vertex}
	\label{Fig:3-vertex}
\end{figure}

As a result of the shift symmetry which is respected by the regulator, $\int
dp f (p_{\mu} ) = \int dp f (p_{\mu} +q_{\mu} )$ for arbitrary $q_{\mu}$, we can
prove the Ward-Takahashi identity:
\begin{equation}
  -i q_{\mu} \Gamma^{\mu} ( p+q,p ) =S^{-1} ( p+q ) -S^{-1} ( p ) \label{wt}
\end{equation}
Here $S ( p ) = \frac{i}{\not{p} -M- \Sigma ( p )}$ is the full electron
propagator and $\Gamma^{\mu} ( p+q,p )$ is the three proper vertex.

Below we explicitly verified that the pole at $d=4$ satisfies (\ref{wt})\footnote{The finite part of the Ward-Takahashi identity is true also, but the computation is too long to show it in this letter.}

Pole contribution:
\begin{eqnarray}
  \mathbbm{P} \Sigma ( q ) =- ( -i e )^{2} \frac{1}{16 \pi^{2}} \left\{ 2m^{2}
  \frac{\not{n}}{n.q} - \not{q} +4M  \right\} \frac{1}{2- \omega} &  & 
  \label{poles}
\end{eqnarray}
\begin{eqnarray}
  \mathbbm{P} \Gamma^{\mu} ( p+q,p ) =- ( -i e )^{2} \frac{1}{16 \pi^{2}}
  \frac{1}{2- \omega} \left( \gamma_{\mu} +2m^{2}   \not{n} \frac{n_{\mu}}{n.p
  n. ( p+q )} \right) &  &  \label{poleg}
\end{eqnarray}
The divergent piece satisfies the Ward identity:
\begin{eqnarray*}
  q_{\mu} \mathbbm{P} \Gamma^{\mu} ( p+q,p ) =\mathbbm{P} \Sigma ( p )
  -\mathbbm{P} \Sigma ( p+q ) =- ( -i e )^{2} \frac{1}{16 \pi^{2}} \frac{1}{2-
  \omega} \left\{ \not{q} -2m^{2} \not{n} \left( \frac{1}{n.p+n.q} -
  \frac{1}{n.p} \right) \right\} &  & 
\end{eqnarray*}

\subsection{Form factors}

The on-shell proper vertex can be written as follows:
\begin{equation}
  \bar{u} ( p+q ) \left\{ G_{2} \left[ -i \sigma_{\mu \nu} q_{\nu} \not{n}
  \right] +G_{3} \not{n} Q_{\mu} +F_{3} \not{n} \sigma_{\mu \nu} q_{\nu}
  \not{n} + \tilde{\gamma}_{\mu} F_{1} +F_{2 } i \frac{\sigma_{\mu \nu}}{2M}
  q_{\nu} \right\} u ( p ) \label{ff}
\end{equation}
where:
\begin{eqnarray*}
  \tilde{\gamma}_{\mu} = \gamma_{\mu} + \frac{m^{2}}{2} \frac{\not{n} 
  n_{\mu}}{n.p  ( n.p+n.q )} , & Q_{\mu} =q_{\mu} -q^{2} \frac{n_{\mu}}{n.q}  
\end{eqnarray*}
$F_{1} ,F_{2} .F_{3} ,G_{2} ,G_{3}$ are forms factors(Lorentz scalar
combinations of $n_{\mu} ,p_{\mu} ,q_{\mu}$). Under the $Sim(2)$
scaling $n_{\mu} \rightarrow \lambda n_{\mu}$, $F_{1} ,F_{2}$ are
invariants,$F_{3} \rightarrow \lambda^{-2} F_{3}$,$G_{3} \rightarrow
\lambda^{-1} G_{3}$,$G_{2} \rightarrow \lambda^{-1} G_{2}$.

In the Non-Relativistic(NR) limit we get Table 1, keeping terms that are at most linear in $q_{\mu}$.
\begin{table}[h]
\begin{center}
	\begin{tabular}{lp{.8\linewidth}}
		NR limit & Form factor\\[5pt]
		$2M_{e} \varphi^{\uparrow}_{s} \varphi_{s} A_{0}$ & $F_{1}$(0)\\
		$\frac{3m^{2}}{4M^{2}} i  \varepsilon_{i j k} \varphi^{\uparrow}_{s}
		\sigma^{i} \varphi_{s'} \hat{n}_{j} q_{k} A_{0}$ & $F_{1}$(0)\\
		$i  \varepsilon_{i j k} q_{j} \varphi^{\uparrow}_{s} \sigma^{k} \varphi_{s'}
		A_{i}$ & $F_{1}$(0)\\
		$\frac{m^2}{2 M^2} (- i \hat{n}_i \varphi^{\uparrow}_s \sigma^a \varphi_{s'}
		\varepsilon_{a b c} \hat{n}_b q_c + i \varepsilon_{i j k} \varphi^{\uparrow}_s
		\sigma^j \varphi_{s'} \hat{n}_k \hat{n}. \vec{q}) A_i$ & $F_{1}$(0)\\
		$-2 i n_{0}  M \varepsilon_{i j k} \varphi^{\uparrow}_{s} \sigma^{k}
		\varphi_{s'} q_{j} A_{i}$ & $G_{2} ( 0 )$\\
		$-i  \varepsilon_{i j k} n_{k} \frac{m^{2}}{M} \varphi^{\uparrow}_{s} \hat{n}
		. \vec{\sigma} \varphi_{s'} q_{j} A_{i}$ & $G_{2} ( 0 )$\\
		$i ( 2M \varepsilon_{i j k}  n_{k} \varphi^{\uparrow}_{s} \sigma^{j}
		\varphi_{s'} +2M_{e}  i n_{i} \varphi^{\uparrow}_{s} \varphi_{s'} ) A_{0} 
		q_{i}$ & $G_{2} ( 0 )$\\
		$2M_{e}  n_{0} \varphi^{\uparrow}_{s} \varphi_{s'} Q_{\mu} A^{\mu}$ & $G_{3} ( 0 )$\\
		$( -4M_{e} \varepsilon_{i j k} n_{k} \varphi^{\uparrow}_{s} \vec{n} .
		\vec{\sigma} \varphi_{s'} +4M_{e} n_{0}^{2} \varepsilon_{i j k}
		\varphi^{\uparrow}_{s} \sigma^{k} \varphi_{s'} ) q_{j} A_{i}$ & $F_{3} ( 0 )$\\
		$4M_{e}  n_{0} \varepsilon_{i j k} n_{j} \varphi^{\uparrow}_{s} \sigma^{k}
		\varphi_{s'} A_{0} q_{i}$ & $F_{3} ( 0 )$\\
		$i  \varepsilon_{i j k} \varphi^{\uparrow}_{s} \sigma^{k} \varphi_{s'} A_{i}
		q_{j}$ & $F_{2} ( 0 )$\\
		$-i \frac{m^{2}}{2M^{2}} \varepsilon_{i j k} \hat{n}_{j}
		\varphi^{\uparrow}_{s} \sigma^{k} \varphi_{s} A_{0} q_{i}$ & $F_{2} ( 0 )$
	\end{tabular}
\end{center}  
  \caption{In the right column we list  the form factor. In the left column
  we have the NR limit of the matrix element accompanying the form factor in
  (\ref{ff}).All form factors are evaluated at $q_{\mu} =0$. Here $A_{0}$ is the electric
  potential and $A_{i}$ is the vector potential.$\varphi_{s'}$ is  a two
  dimensional constant vector that corresponds to the NR limit of the Dirac
  spinors.
}
\end{table}

To show the power of the $Sim(2)$ invariant regularization
prescription presented in this letter, we will compute the one loop
contribution to the (isotropic)anomalous magnetic moment of the electron. It
is given by $F_{2} ( 0 ) -2n_{0} M G_{2} ( 0 ) -4F_{3} ( 0 ) M_{e} n_{0}^{2}
i$(See rows 11, 5 \ and 9 \ of Table 1).

Introduce the following integrals:
\begin{eqnarray*}
  \int d k  ( n.k+n.q )^{a1} ( n.k )^{a2} ( ( k-p )^{2} )^{a3} ( k^{2}
  -M_{e}^{2} )^{a4} ( ( k+q )^{2} -M_{e}^{2} )^{a5} =I ( a1,a2,a3,a4,a5 ) &  &
  \\
  \int d k  ( n.k+n.q )^{a1} ( n.k )^{a2} ( ( k-p )^{2} )^{a3} ( k^{2}
  -M_{e}^{2} )^{a4} ( ( k+q )^{2} -M_{e}^{2} )^{a5} k_{\mu} = &  & \\
  I_{11} ( a1,a2,a3,a4,a5 ) p_{\mu} +I_{12} ( a1,a2,a3,a4,a5 ) q_{\mu} +I_{13}
  ( a1,a2,a3,a4,a5 ) n_{\mu} &  & \\
  \int d k  ( n.k+n.q )^{a1} ( n.k )^{a2} ( ( k-p )^{2} )^{a3} ( k^{2}
  -M_{e}^{2} )^{a4} ( ( k+q )^{2} -M_{e}^{2} )^{a5} k_{\mu} k_{\nu} = &  & \\
  I_{21} ( a1,a2,a3,a4,a5 ) \eta_{\mu \nu} +I_{22} ( a1,a2,a3,a4,a5 ) p_{\mu}
  p_{\nu} +I_{23} ( a1,a2,a3,a4,a5 ) q_{\mu} q_{\nu} + &  & \\
  I_{24} ( a1,a2,a3,a4,a5 ) n_{\mu} n_{\nu} +I_{25} ( a1,a2,a3,a4,a5 ) (
  p_{\mu} q_{\nu} +p_{\nu} q_{\mu} ) + &  & \\
  I_{26} ( a1,a2,a3,a4,a5 ) ( p_{\mu} n_{\nu} +p_{\nu} n_{\mu} ) +I_{27} (
  a1,a2,a3,a4,a5 ) ( n_{\mu} q_{\nu} +n_{\nu} q_{\mu} ) &  & 
\end{eqnarray*}
We get:
\begin{eqnarray*}
  F_{2} ( 0 ) -2n_{0} M G_{2} ( 0 ) -4F_{3} ( 0 ) M_{e} n_{0}^{2} i= &  & \\
  -4i e^{2} M^{2} \{ I_{22} ( 0,0,-1,-1,-1 ) -I_{11} ( 0,0,-1,-1,-1 ) \} - & 
  & \\
  2i e^{2} m^{2} \{ -I ( 0,0,-1,-1,-1 ) -2I_{22} ( 0,-1,-1,-1,-1 ) M n_{0}
  +3I_{11} ( 0,0,-1,-1,-1 ) \} &  & 
\end{eqnarray*}
Evaluating the integrals according to the $Sim(2)$ invariant
prescription to $o ( m^{2} )$, we get:
\begin{eqnarray*}
  F_{2} -4F_{3} M_{e} n_{0}^{2} i-2G_{2} n_{0} M= \frac{\alpha}{2 \pi} &  & 
\end{eqnarray*}
where $\alpha$ is the fine structure constant. Therefore to this order the QED
result holds.

Notice that already at tree level, the model predicts the existence of an
anisotropic electric  moment of the electron, corresponding to the second
line of the list and an anisotropic magnetic moment of the electron,
corresponding to the fourth row of the list, both of  the order of
$\frac{m^{2}}{M_{e}^{2}}$. The electric dipole moment is:
\[ | \vec{p} | = \frac{3e}{4M_{e}} \frac{m^{2}}{M_{e}^{2}} | ( \vec{s} \times
   \hat{n} ) | \lneq \frac{3}{8} \lambda e \frac{m^{2}}{M_{e}^{2}} \]
where $\lambda =2.4 \times 10^{-12} m$ is the Compton wave length of the
electron.

Using the best bound on the electric dipole moment of the
electron{\cite{acme}}, $| \vec{p} | < 8.7 \times 10^{-29} ~ e\dot cm.$,we get:
\[ \frac{m^{2}}{M_{e}^{2}} <9.7 \times 10^{-19} \]
For the muon $\lambda =1.17 \times 10^{-14} m$. Using the best bound on the
muon electric dipole moment{\cite{bennett}},$| \vec{p}_{\mu} | < 1.8 \times
10^{-19} ~ e\dot cm.$,we get:
\[ \frac{m_{\mu}^{2}}{M_{\mu}^{2}} <4 \times 10^{-7} \]
Bounds using the experimental values of the magnetic moments are much weaker\cite{dunn}.

The $Sim(2)$ invariant regularization opens the way to explore the
full quantum possibilities of VSR. They should be systematically studied, in
Particle Physics models as well as in Quantum Gravity models.

{Acknowledgements}

The work of J. A. is partially supported by Fondecyt 1150390, Anillo ACT 1417.
J.A. wants to thank H. Morales-T{\'e}cotl, L.F. Urrutia, D. Espriu \ and \ R.
Soldati for useful remarks.

\end{document}